\documentclass[dvips]{article}
\usepackage{supertabular,lscape,epsfig}
\usepackage{amssymb}
\usepackage{amsmath}

\DeclareSymbolFont{ppa}{OT1}{ppl}{m}{it}
\DeclareMathSymbol{\vv}{\mathalpha}{ppa}{'166}

\begin{document}

\newcommand{\dd}{\,{\rm d}}
\newcommand{\ie}{{\it i.e.},\,}
\newcommand{\etal}{{\it et al.\ }}
\newcommand{\eg}{{\it e.g.},\,}
\newcommand{\cf}{{\it cf.\ }}
\newcommand{\vs}{{\it vs.\ }}
\newcommand{\zdot}{\makebox[0pt][l]{.}}
\newcommand{\up}[1]{\ifmmode^{\rm #1}\else$^{\rm #1}$\fi}
\newcommand{\dn}[1]{\ifmmode_{\rm #1}\else$_{\rm #1}$\fi}
\newcommand{\upd}{\up{d}}
\newcommand{\uph}{\up{h}}
\newcommand{\upm}{\up{m}}
\newcommand{\ups}{\up{s}}
\newcommand{\arcd}{\ifmmode^{\circ}\else$^{\circ}$\fi}
\newcommand{\arcm}{\ifmmode{'}\else$'$\fi}
\newcommand{\arcs}{\ifmmode{''}\else$''$\fi}
\newcommand{\MS}{{\rm M}\ifmmode_{\odot}\else$_{\odot}$\fi}
\newcommand{\RS}{{\rm R}\ifmmode_{\odot}\else$_{\odot}$\fi}
\newcommand{\LS}{{\rm L}\ifmmode_{\odot}\else$_{\odot}$\fi}

\newcommand{\Abstract}[2]{{\footnotesize\begin{center}ABSTRACT\end{center}
\vspace{1mm}\par#1\par
\noindent
{~}{\it #2}}}

\newcommand{\TabCap}[2]{\begin{center}\parbox[t]{#1}{\begin{center}
  \small {\spaceskip 2pt plus 1pt minus 1pt T a b l e}
  \refstepcounter{table}\thetable \\[2mm]
  \footnotesize #2 \end{center}}\end{center}}

\newcommand{\TableSep}[2]{\begin{table}[p]\vspace{#1}
\TabCap{#2}\end{table}}

\newcommand{\FigCap}[1]{\footnotesize\par\noindent Fig.\  %
  \refstepcounter{figure}\thefigure. #1\par}

\newcommand{\TableFont}{\footnotesize}
\newcommand{\TableFontIt}{\ttit}
\newcommand{\SetTableFont}[1]{\renewcommand{\TableFont}{#1}}

\newcommand{\MakeTable}[4]{\begin{table}[htb]\TabCap{#2}{#3}
  \begin{center} \TableFont \begin{tabular}{#1} #4 
  \end{tabular}\end{center}\end{table}}

\newcommand{\MakeTableSep}[4]{\begin{table}[p]\TabCap{#2}{#3}
  \begin{center} \TableFont \begin{tabular}{#1} #4 
  \end{tabular}\end{center}\end{table}}

\newenvironment{references}%
{
\footnotesize \frenchspacing
\renewcommand{\thesection}{}
\renewcommand{\in}{{\rm in }}
\renewcommand{\AA}{Astron.\ Astrophys.}
\newcommand{\AAS}{Astron.~Astrophys.~Suppl.~Ser.}
\newcommand{\ApJ}{Astrophys.\ J.}
\newcommand{\ApJS}{Astrophys.\ J.~Suppl.~Ser.}
\newcommand{\ApJL}{Astrophys.\ J.~Letters}
\newcommand{\AJ}{Astron.\ J.}
\newcommand{\IBVS}{IBVS}
\newcommand{\PASP}{P.A.S.P.}
\newcommand{\Acta}{Acta Astron.}
\newcommand{\MNRAS}{MNRAS}
\renewcommand{\and}{{\rm and }}
\section{{\rm REFERENCES}}
\sloppy \hyphenpenalty10000
\begin{list}{}{\leftmargin1cm\listparindent-1cm
\itemindent\listparindent\parsep0pt\itemsep0pt}}%
{\end{list}\vspace{2mm}}

\def\TYLDA{~}
\newlength{\DW}
\settowidth{\DW}{0}
\newcommand{\dw}{\hspace{\DW}}

\newcommand{\refitem}[5]{\item[]{#1} #2%
\def\REFARG{#3}\ifx\REFARG\TYLDA\else, {\it#3}\fi
\def\REFARG{#4}\ifx\REFARG\TYLDA\else, {\bf#4}\fi
\def\REFARG{#5}\ifx\REFARG\TYLDA\else, {#5}\fi.}

\newcommand{\Section}[1]{\section{\hskip-6mm.\hskip3mm#1}}
\newcommand{\Subsection}[1]{\subsection{#1}}
\newcommand{\Acknow}[1]{\par\vspace{5mm}{\bf Acknowledgements.} #1}
\pagestyle{myheadings}

\newfont{\bb}{ptmbi8t at 12pt}
\newcommand{\xrule}{\rule{0pt}{2.5ex}}
\newcommand{\xxrule}{\rule[-1.8ex]{0pt}{4.5ex}}
\def\thefootnote{\fnsymbol{footnote}}
\begin{center}
{\Large\bf Evolutionary Status of Late-Type Contact Binaries}
\vskip.6cm
{\bf K.~~ S~t~{\c e}~p~i~e~{\'n}}
\vskip2mm
Warsaw University Observatory, Al.~Ujazdowskie~4, 00-478~Warszawa, Poland\\
e-mail: kst@astrouw.edu.pl\\
\end{center}
\vskip3mm
\centerline{\it Received March 28, 2006}
\vskip3mm

\Abstract{The old model of an unevolved, cool contact binary, in which the
secondary component is strongly oversized due to energy transfer from the
primary, and the whole system is out of thermal equilibrium, encounters
serious problems.

I present a new scenario for evolution of contact binaries, which solves
the problem of thermal nonequilibrium by assuming that contact binaries are
past mass exchange with a mass ratio reversal. The scenario is divided
into three phases. In Phase~I loss of angular momentum (AM) due to
magnetized wind of a detached binary is followed until the primary
component fills its critical Roche lobe. In Phase~II mass transfer takes
place until mass ratio reversal. Arguments are given in favor of such a
process in pre-contact binaries. In Phase~III an approximate evolutionary
path of the contact binary is followed until a possible coalescence. AM
loss, evolutionary effects of the components and mass transfer to the
primary are taken into account.

It is concluded that W~UMa type binaries are old objects with secondaries
in an advanced evolutionary stage, possibly with small helium cores. Both
components fulfill the mass-radius relation for contact binaries while
being in thermal equilibrium.}{Stars: activity -- binaries: close -- Stars:
evolution -- Stars: late-type -- Stars: rotation}

\Section{Introduction}				     
W~UMa type stars are binaries of spectral type F0--K5 which have both
components surrounded by a common envelope lying between the inner and
outer Lagrangian zero velocity equipotential surfaces (Mochnacki 1981).

Kuiper (1941) noted that contact binaries with unequal zero-age
components should not exist because radii of the components must fulfill
two, mutually contradictory conditions: one resulting from the
mass-radius relation for zero-age stars and the other, relating sizes of
the Roche lobes, identical in this case with stellar sizes, to stellar
masses. The fact that contact binaries are nevertheless observed is
known as ``Kuiper paradox''. Lucy (1968) stressed that such a
configuration can be either the result of subsequent evolution of
zero-age detached systems or it is not in equilibrium. Following Kuiper
(1941) he assumed the latter to be true and developed a theory of
zero-age contact binaries with a common convective envelope. The theory
explained the observed light curves of W~UMa type binaries but
encountered several problems when confronted with other observations of
these stars (\eg Mochnacki 1981, 1985, Rucinski 1993). On the other
hand, several observational and theoretical facts point toward the
advanced evolutionary status of W~UMa type binaries. Numerical
simulations of a binary formation favor an early fragmentation of a
protostellar cloud resulting in a formation of a detached binary (Boss
1993). Fission of a rapidly rotating protostar into two cores was never
observed (Bonnel 2001). This result is in agreement with a lack of
contact binaries among T~Tau stars and among members of the youngest
stellar clusters. Unless the exceptional orbital angular momentum loss
(AML) takes place, the shortest period zero-age binaries with solar type
components should have periods about 2~days or more (see the discussion
by St{\c e}pie{\'n} 1995). This is several times more than required for a
contact configuration. However, stars of the lower main sequence (MS)
lose angular momentum (AM) {\it via} a magnetized wind with efficiency
increasing with increasing angular velocity (Mestel 1984, Kawaler 1988,
St{\c e}pie{\'n} 1991). Assuming synchronous rotation of the close binary
components AML results in a decrease of orbital AM and shortening of the
orbital period. Stars approach each other and ultimately form a contact
system (Vilhu 1982, Mochnacki 1985). Closer consideration of this
mechanism shows that it takes several Gyr for a detached binary with an
initial period of the order of 2~days to lose enough AM and to form a
contact system (St{\c e}pie{\'n} 1995). For longer periods it takes
correspondingly more time. Observations of W~UMa type stars in old
stellar clusters confirm this result. They are virtually unknown in
young clusters but their frequency of occurrence rapidly increases for
open clusters with age exceeding 4--5~Gyr (Kaluzny and Rucinski 1993,
Rucinski 1998) and for globular clusters (Rucinski 2000). Note, however,
that the AML time scale of $\approx5{-}14$~Gyr is equal to the nuclear
time scale for stars with masses 0.9--1.3~\MS. Because of this
coincidence, massive primaries can evolve up to, or even beyond terminal
age MS (TAMS) by the time their shrinking Roche lobes reach the stellar
surface. Filling up the Roche lobe by the more massive component results
in mass exchange, a mass ratio reversal and the formation of a contact
binary in which a secondary has a hydrogen depleted core.

A relatively high frequency of occurrence of W~UMa type stars among MS
dwarfs, particularly in the Galactic bulge \ie among very old disk stars
(Rucinski 1998, 2002, Szyma\'nski, Kubiak and Udalski 2001) indicates
that the total life time of a contact configuration cannot be as short
as a stellar thermal time scale of, say, 10 to 100 million years, but it
should last several Gyr. This argument can be made more quantitative.
Halbwachs \etal (2004) present the latest statistics concerning the
properties of binary stars of spectral types F7--K. According to their
results there are 56 binaries with orbital periods shorter than 10 years
among 405 stars of the unbiased sample. This gives the frequency of
13.5\%. Of these binaries two (possibly three) have periods shorter than
3 days -- short enough to form a contact binary during their MS
evolution. This gives a frequency of short period binaries equal to
0.005--0.007. Another estimate of this frequency can be obtained assuming
that the observed binary distribution is described by the Duquennoy and
Mayor (1991) formula. Integration of their formula over periods shorter
than 3 days and then over periods shorter than 10 years gives a value of
0.02 for the ratio of both integrals. This multiplied by 13.5\% gives
0.003 for the frequency of short period stars. Both methods give similar
results suggesting that this frequency is in the range of 0.003--0.007.
Frequency of occurrence of W~UMa stars was recently determined anew by
Rucinski (2002). According to him it is equal to 0.002 for the local
field stars and to 0.008 for the OGLE~I sample. Approximate equality of
both frequencies, \ie of short period detached binaries and contact
binaries indicates that the life time of a binary in the contact phase
is of the same order as the life time in the detached phase. The latter
time is equal to several Gyr as it was argued above.

To sum up the above arguments we see that a typical W~UMa type binary
enters the contact configuration when it is at least 4--5 Gyr old and spends
several more Gyr in this phase. This conclusion agrees with the age of 8
Gyr obtained for the field W~UMa stars from their space motions (Guinan and
Bradstreet 1988). In the light of these results the basic assumption of
Lucy (1968) that W~UMa type stars are very young turns out to be incorrect.

It will be argued in the present paper that a typical W~UMa star is a
binary past mass exchange in the case A or early B (\ie when the
initially more massive component is close to, or immediately past TAMS).
After mass ratio reversal the presently more massive component is a
weakly evolved MS star whereas the less massive component is an
evolutionary advanced star with hydrogen depleted in its center. Such a
pair of stars can fulfill the mass-radius relation for contact binaries
while being in thermal equilibrium. This solves the Kuiper paradox in a
natural way.

The paper is organized as follows: in Section~2 observational data on radii
and masses of components of W~UMa stars are compared with evolutionary
model calculations of single stars. Section~3 lists basic assumptions and
equations used to describe the evolution of a binary. Next, the evolution
of the orbital parameters of one binary with a moderate mass ratio and the
total mass of 1.8~\MS is discussed. Three main phases are distinguished:\\
I -- approach to contact,\\
II -- mass exchange on a thermal time scale,\\
III -- evolution of a contact binary until merging of the components.

The last Section contains the main conclusions of the paper.

\Section{Comparison of Observations with Evolutionary Models of Single Stars}
The most complete, recent set of physical parameters of contact binaries,
obtained in a uniform way, has been published by Maceroni and van't Veer
(1996). Their Table~3 contains absolute parameters of 78 binaries, which
were calculated using the available at that time photometric and
spectroscopic data. Due to lack of spectroscopic data for many stars,
photometric mass ratios have been used which increases uncertainty of the
listed parameters for these stars. Later, spectroscopic mass ratios have
been determined for several such stars. In some cases their values agree
well with the photometric values but in other cases, like AO Cam and LS~Del
they badly disagree. Such stars were ignored when plotting Fig.~1.
\begin{figure}[htb]
\centerline{\includegraphics[width=12.5cm]{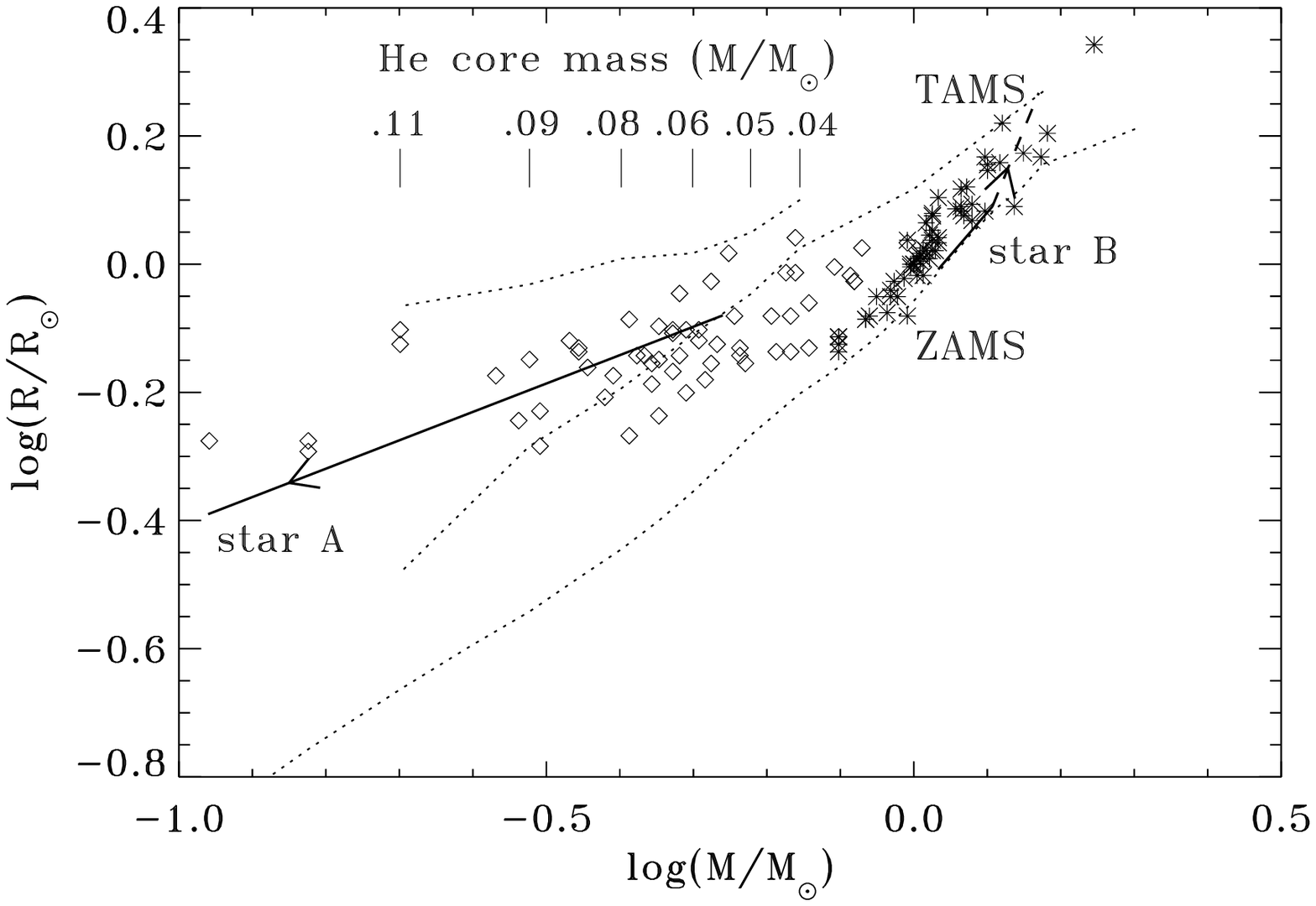}}
\FigCap{Observed radii of the components of W~UMa type binaries \vs
mass. Asterisks correspond to primaries, \ie the presently more massive
components and diamonds to secondaries. Dotted lines give, respectively,
the mass-radius relations for ZAMS, TAMS and for evolved stars
possessing helium cores with masses as indicated by the numbers along
the line. Heavy straight lines with arrows, labeled ``star A'' and
``star B'' represent schematic evolutionary tracks of both components of
the discussed binary during Phase~III (see Section~3.4).}
\end{figure}

We restrict our attention to cool contact binaries with $M_{\rm
tot}\le2~\MS$, where $M_{\rm tot}$ is the total mass of the binary (see
Maceroni and van't Veer 1996). All binaries except AH~Cnc, classified as
W-type by Maceroni and van't Veer, are included into our analysis but
several massive A-type binaries are not. The data given by Maceroni and
van't Veer (1996) are plotted in Fig.~1. The primary components which we
denote B-components (see Section~3) are marked with asterisks and the
secondary, less massive A-components are marked with diamonds. Dotted lines
give mass-radius relations for (from bottom to top): ZAMS stars, TAMS stars
and evolved stars with small helium cores. They are based on models
obtained by VandenBerg (1985, see also Rucinski 1992), Schaller \etal
(1992) and Sienkiewicz.\footnote{At a request of the present author a set
of evolutionary models of single, low mass stars was calculated by Dr.\
Ryszard Sienkiewicz using his recently updated evolution code. Models with
masses 0.1--0.7~\MS were evolved (at a constant mass) from ZAMS up to the
formation of helium core of 0.08--0.15 of the solar mass (depending on the
stellar mass).} Masses of the helium cores were selected in such a way that
the resulting mass-radius relation for the evolved stars forms an upper
bound to the observed values.

Inspection of Fig.~1 confirms the conclusion reached by Hilditch \etal
(1988) that all primaries of the W-type binaries lie at, or close to ZAMS
whereas their secondaries lie close to, or beyond TAMS. In case of A-type
binaries with extreme mass ratios primaries approach TAMS in the
mass-radius diagram although they lie closer to ZAMS in the mass-luminosity
diagram (Yakut and Eggleton 2005), whereas secondaries lie significantly
above TAMS, with radii bound from above by models with small helium
cores. Such a situation when a less massive component of a binary is more
advanced evolutionary, is known as an Algol paradox and is an indication
that the system was subject to the mass exchange process resulting in a
reversal of the mass ratio (Paczy\' nski 1971). It is assumed here that
the same process has taken place in most of the observed W~UMa stars. As a
result, the presently less massive components occupy the position at, and
above TAMS (Fig.~1) because they are oversized due to their advanced
evolutionary stage and not due to energy transfer from the more massive
components.

\Section{Evolutionary Scenario -- a Binary with Different Mass Components}
A great variety of initial conditions of close binary stars, in particular
of masses of individual components and orbital periods results in a
plethora of their possible evolutionary scenarios (\eg Eggleton 1996,
Eggleton and Kiseleva-Eggleton 2002, Yakut and Eggleton 2005). It is beyond
a scope of the present paper to determine precisely a range of the initial
parameters of binaries leading to W~UMa type stars. Instead, evolution of
orbital parameters of one selected binary from ZAMS till merger of the
components will be followed in detail and discussed (preliminary results
are given by St{\c e}pie{\'n} 2004).

The following main phases will be considered:\\
I) AML due to magnetized wind during the MS evolution and filling of
the Roche lobe by the more massive A component,\\
II) fast mass exchange on a thermal time scale till reversal of the
mass ratio and formation of a contact configuration,\\
III) slow evolution of the contact binary toward the extreme mass
ratio and ultimate merging of the components.

\subsection{Initial Configuration}
As an initial configuration a detached ZAMS binary with masses $M_{\rm
A,i}{=}1.2~\MS$, $M_{\rm B,i}=0.6~\MS$ is adopted. The mass of the larger
component is close to the upper mass limit for stars possessing strong
coronal emission associated with a significant mass loss {\it via}
magnetized wind. The emission decreases rapidly for stars with still larger
masses (Schmitt \etal 1985). The mass of the smaller component was assumed
to be significantly lower than that of the larger component to stress the
difference in evolutionary effects of both components on the resulting
contact configuration and to obtain a ``typical'' W~UMa system with mass
ratio close to 0.5 (Shu and Lubow 1981) after the mass exchange. Binaries
with the initial mass ratio closer to one will be discussed later.

The initial mass ratio $q_{\rm i} =M_{\rm A,i}/M_{\rm B,i}=2$ (note
that $q$ is defined throughout the paper as the ratio of the mass of
component A to the mass of component B) whereas the initial total mass of
the binary is $M_{\rm tot,i}=M_{\rm A,i}+M_{\rm B,i}=1.8~\MS$. The
assumed value of the initial orbital period, $P_{\rm orb,i}=2$ days, is
equal to the short limit period of cool ZAMS binaries adopted by St{\c e}pie{\'n}
(1995). The radii of both components, semi-axis $a$ of the orbit and the
initial AM are given in Table~1.

It is assumed in the following that both components are active and lose AM
{\it via} magnetized wind and that they rotate in a full synchronization
with the orbital period. Their spin AM is neglected compared to orbital AM.

\MakeTable{lccccccccc}{12.5cm}{Evolutionary history of the discussed binary}
{\hline
\noalign{\vskip3pt}
evolutionary & age & $M_{\rm A}$ & $M_{\rm B}$ & $q$ & $R_{\rm
A}$ & $R_{\rm B}$ & $P_{\rm orb }$ & $a$ & $H_{\rm orb}$ \\
stage & Gyr & ~\MS &~\MS & &~\RS &~\RS & days &
\RS & $10^{51}~{\rm g\cdot cm^2/s}$ \\
\noalign{\vskip3pt}
\hline
\noalign{\vskip3pt}
ZAMS & 0 & 1.2 & 0.6 & 2 & 1.1 & 0.49 & 2 & 8 & 9.2 \\
End of Phase I & 6 & 1.14 & 0.54 & 2.1 & 1.3 & 0.5 & 0.46 & 3 & 5.1 \\
End of Phase II & 6.1 & 0.55 & 1.08 & 0.51 & 0.8 & 1.02 & 0.32 & 2.3 & 4.3 \\
Coalescence & 11 & 0.1 & 1.48 & 0.07 & 0.4 & 1.3 & 0.46 & 2.2 & 1.1 \\
\noalign{\vskip3pt}
\hline}

\subsection{Phase I}
Loss of AM momentum of a cool close binary {\it via} magnetized wind was
discussed by Vilhu (1982) and, more recently, by St{\c e}pie{\'n} (1995). The
latter author derived and calibrated a semi-empirical formula for the
orbital period variation of a close cool binary, based on the rotation
evolution of solar type stars. The formula does not contain any free
parameters.
$$\frac{{\rm d}\,P_{\rm orb}}{{\rm d}\,t}=-(2.6\pm1.3)
\times10^{-10}P_{\rm orb}^{-1/3}{\rm e}^{-0.2P_{\rm orb}}\eqno(1)$$
where $P_{\rm orb}$ is in days and time in years. For very short orbital
periods the exponential factor is close to unity and it varies very little
during the subsequent evolution of the orbital period of the discussed
binary (in our case, from 2 days to about 0.3 of a day). It can therefore
be left out. This simplifies significantly the following
considerations. Note, that the 50\% uncertainty of the numerical factor on
RHS of Eq.~(1) is much larger than the error resulting from such an
approximation.

The third Kepler law can be expressed as 
$$P_{\rm orb}=0.1159\sqrt{a^3/M_{\rm tot}}\eqno(2)$$
where $a$, the semi-axis, is in units of solar radius and $P_{\rm orb}$
again in days. 

From these two formulas an equation for the time derivative of 
semi-axis $a$ can be derived and integrated
$$a=\sqrt{a^2_{\rm init}-9\times10^{-9}t}\eqno(3)$$
where the initial value of semi-axis $a_{\rm init}$ is in solar
radii and time $t$ is in years. The uncertainty of the numerical
coefficient in Eq.~(3) is of the same order as the one in Eq.~(1). 

Effective radii of the Roche lobes of both components are approximated
to better than 1\% by formulae (Eggleton 1983)
$$\frac{r_{\rm A}}{a}=\frac{0.49q^{2/3}}{0.6q^{2/3}+\ln{(1+q^{1/3})}},\eqno(4)$$
$$\frac{r_{\rm B}}{a}=\frac{0.49q^{-2/3}}{0.6q^{-2/3}+\ln{(1+
q^{-1/3})}}.\eqno(5)$$

Our initial binary has semi-axis $a_{\rm init}\approx8~\RS$ (Table~1).
It follows from Eq.~(3) that after about 6~Gyr it decreases to
$a\approx~3~\RS$. The size of the Roche lobe of star A is then $r_{\rm
A}=0.44a\approx1.3~\RS$. Approximately at the same time star A reaches
TAMS and its radius increases to the same value of 1.3~\RS so the star
fills up its Roche lobe (Fig.~2 top). The size of the critical lobe of
star B is at that time $r_{\rm B}=0.32a=0.96~\RS$, \ie it is nearly
twice as large as the stellar radius which has practically not changed
since ZAMS (Fig.~2 bottom).
\begin{figure}[htb]
\centerline{\includegraphics[width=12.5cm]{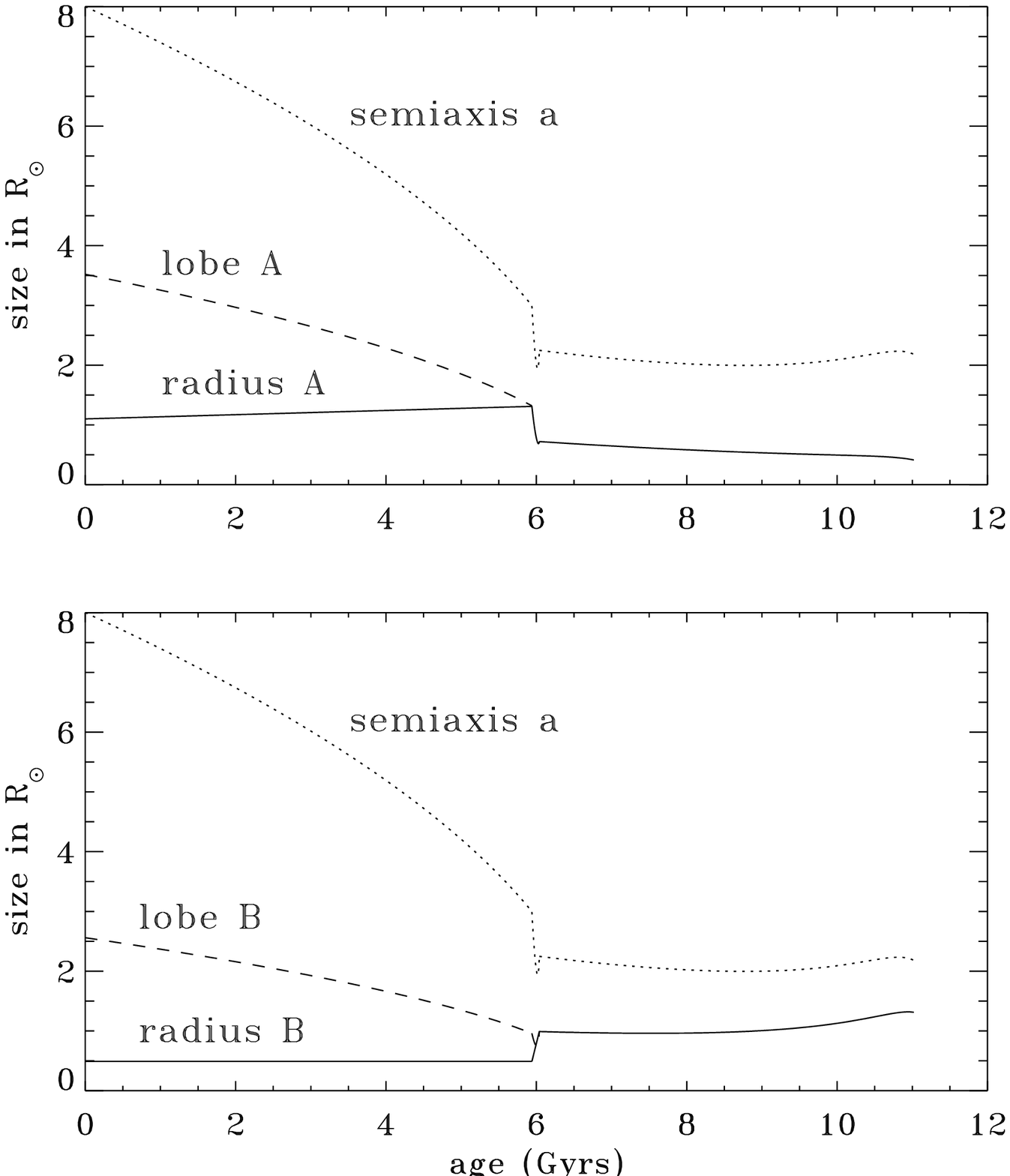}}
\vskip6pt
\FigCap{Variations in time of the radius of star A ({\it top}) and star
B ({\it bottom}) compared with the variations of the orbital semi-axis
and size of the respective critical Roche lobe over all three phases of
evolution discussed in the text. The curves are labeled correspondingly.
Phase~I (left-hand part of each figure) is separated from Phase~III
(right-hand part) by a narrow dip in variation of semi-axis, accompanied
by rapid changes of radii of both stars, which then merge with sizes of
the corresponding Roche lobes (Phase~II).}
\end{figure}

The wind carries away not only AM but also mass. Observations show that the
mass loss rate due to magnetized wind in the most active single stars is of
the order of $10^{-11}$~\MS/year (Wood \etal 2002). During 6~Gyr such a
star loses ${\approx0.06}$~\MS. Assuming that both components of the binary
lose mass at the same rate, their masses become equal to 1.14~\MS and
0.54~\MS, respectively, at the end of Phase~I.

The value of the orbital AM can be computed from the formula
$$H^2_{\rm orb}=GM^3_{\rm tot}\frac{aq^2}{(1+q)^4}.\eqno(6)$$
After 6~Gyr the orbital AM decreases to about 55\% of its initial value.

\subsection{Phase II}
When star A fills its Roche lobe mass transfer begins. The general
paradigm says that star B will react to the mass flux by rapid swelling.
After $\approx0.1$~\MS has been transferred, star B fills its Roche
lobe and a contact binary is formed (Sarna and Fedorova 1989 and
references therein). At this point, convective envelopes of both
components merge with a single value of specific entropy (Lucy 1968). 
The upraised value of the entropy in the convective envelope of star B
(compared to its value in the detached phase) keeps it oversized. Any
additional mass transfer is blocked but energy flows from star A to star
B through the neck between the stars. The resulting configuration is out
of thermal equilibrium and oscillates around the marginal contact (Lucy
1976, Flannery 1976). If both stars are unevolved, the faster nuclear
evolution of star A results in net mass flow from star B and its
ultimate swallowing by star A as it ascends the red giant branch. To sum
up, after the initial, modest mass transfer from star A, star B swells
irreversibly and it transfers all its mass to star A on a nuclear time
scale of the accreting star, until merger occurs.

Is the paradigm correct? One of its basic assumptions, that contact
binaries are unevolved objects, is certainly wrong. Observations and
theoretical arguments show that W~UMa stars are formed from initially
detached binaries and that this process takes at least several Gyr. It
means that primaries with masses close to, or exceeding 1~\MS have enough
time to approach, or even to pass beyond TAMS. Unless the mass ratio is
close to unity, both stars have therefore very different internal
structures when mass transfer begins. The binary resembles a pre-Algol
configuration. Classical Algols originate from binaries with more massive
primaries, not possessing subphotospheric convective zones, and with higher
orbital AM, compared to W~UMa type stars. They emerge from a common
envelope phase (assumed to be formed during a rapid mass transfer) with
mass ratio reversed.

The assumption that convective envelopes of both components of a cool
contact binary must lie on the same adiabatic curve, has been challenged
by Shu \etal (1979) but their hypothesis of a contact discontinuity
requires the energy flow from cooler to hotter medium. Recently,
however, new models of energy flow between components of a cool contact
binary have been developed in which the convective zone of a less
massive component is separated from the common envelope lying above the
critical Roche surface (Martin and Davey 1995, K\"ahler 2002a, 2002b,
2004). The models do not violate the second law of thermodynamics but
they require another mechanism for adjusting the sizes of the components
to the Roche geometry. While such models indicate only a possibility of
solving the problem of cool contact binaries in an alternative way, the
conclusive evidence for the existence of cool close binaries with mass
ratio reversed came from observations. Recent, more accurate modeling of
several contact, or nearly contact binaries resulted in detection of a
significant number of Algols with periods shorter than 1~day (Rucinski
1993). Many of them have orbital periods of 0.5--0.6 of a day (Shaw
1994, Pribulla, Kreiner and Tremko 2003, Budding \etal 2004). An Algol
with the shortest known period is W~Crv with $P_{\rm orb}=0.38$ of a day
(Rucinski and Lu 2000). It is clear that the paradigm does not apply to
these systems and their existence proves that we are still far from
satisfactory understanding of the mass transfer process in close binary
stars (see also Shu and Lubow 1981). Note that all the very short period
Algols will soon become contact binaries due to AML and evolutionary
radius increase of the presently more massive components. It seems
therefore fully justified to abandon the requirement of equal entropy
convective zones and to assume that a typical W~UMa type star has gone
through mass transfer with a mass ratio reversal just as Algols do. The
possibility of existence of contact binaries with mass ratio reversed
has already been invoked in the past by several authors discussing
individual systems with special properties (Tapia and Whelan 1975,
Kraicheva, Tutukov and Yungelson 1986, Sarna and Fedorova 1989, Eggleton
1996) but these systems were considered to be exceptions from the
general rule stating that contact binaries have not reversed mass
ratios. An opposite view is accepted in the present paper. It assumes
that few systems are in the phase discussed \eg by Sarna and Fedorova
(1989) in which the net mass transfer proceeds from a more to a less
massive component but the great majority of the cool contact, or near
contact systems is past the mass ratio reversal. A similar view was
expressed by Iben and Livio (1993) who suggest that secondaries of W~UMa
type binaries should be more advanced evolutionary than primaries.

The following scenario for Phase II is adopted. When star A overflows
its Roche lobe it starts transferring mass to star B. Whatever the
reaction of star B is, we assume that it can accept a significant
portion of mass on a short (thermal in this case) time scale. Apart from
not fully understood details of the evolution of the system in the
common envelope, we assume that the mass transfer continues until the
equilibrium radius of star A becomes smaller than its Roche lobe (a
standard assumption when considering formation of Algols). This occurs
after the mass ratio reversal. Fig.~3 shows the process as a function of
$q$. The mass exchange begins when $q\approx2.1$ and it ends when the
equilibrium radius of star A does not exceed the Roche lobe. Fig.~3 does
not describe an exact behavior of stellar radii during the mass exchange
phase. The curves labeled ``equil. rad. A'' and ``equil. rad. B'' are
used only to determine the beginning and the end of the mass exchange
episode. They are equal to the TAMS (equil. rad. A) and ZAMS (equil.
rad. B) radii of stars with masses varying during the exchange.
\begin{figure}[htb]
\centerline{\includegraphics[width=12.1cm]{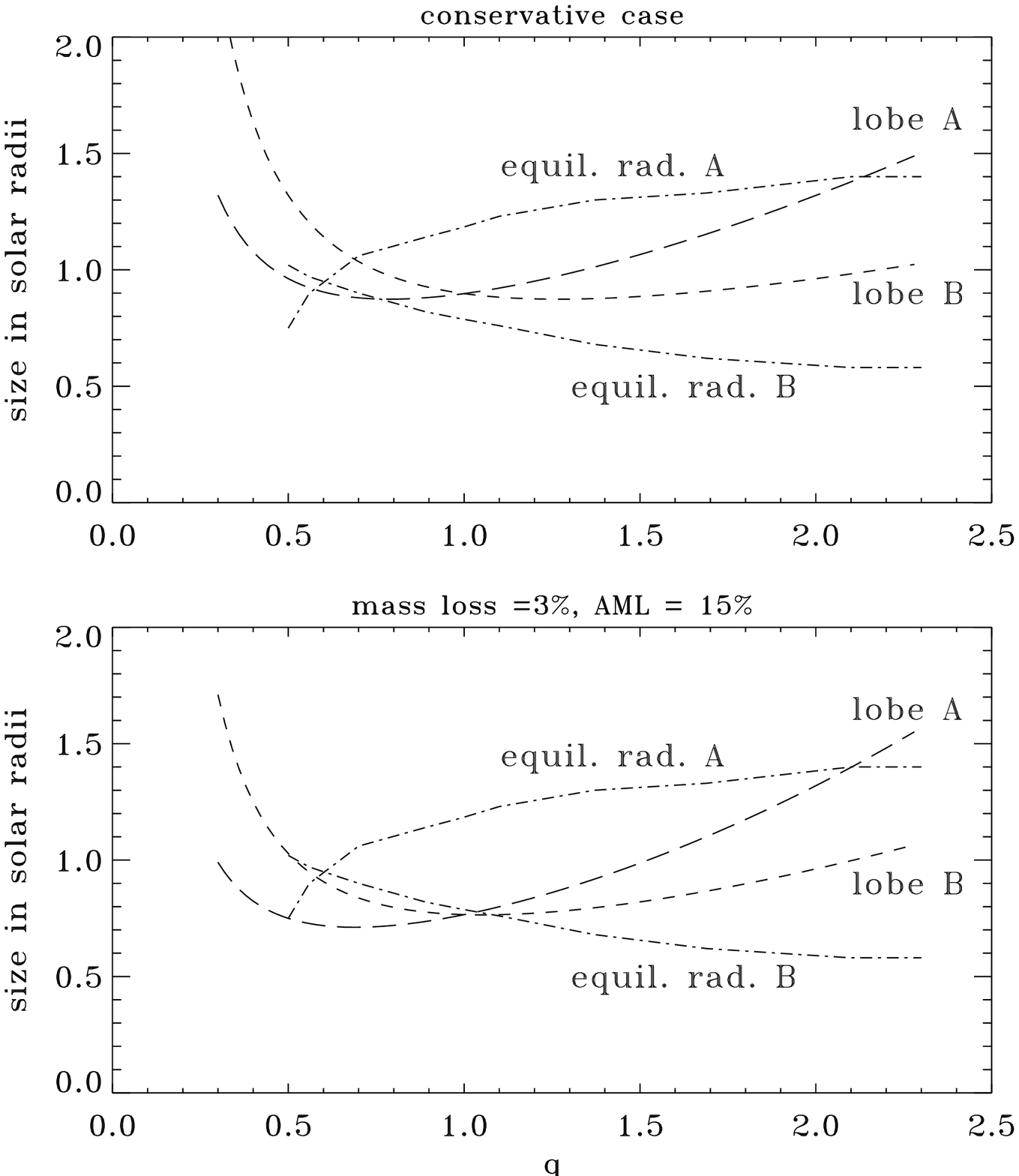}}
\vskip6pt
\FigCap{Variations of the critical Roche lobe sizes (dashed lines)
during the fast mass exchange (Phase~II). Mass exchange begins at
$q\approx2.1$ when the equilibrium radius of star A reaches the size of
its Roche lobe and ends around $q\approx0.5$ when it decreases below the
lobe size. During the mass transfer the equilibrium radii of both stars
(dash-dotted lines) do not describe their actual radii because stars
develop a common envelope, very likely reaching beyond an outer critical
surface. In a conservative case ({\it top}) the equilibrium radius of
star B is at the end of mass transfer smaller than the size of its Roche
lobe which results in a formation of a short-period Algol. Loss of
$\approx15\%$ of the orbital AM during the common envelope phase results
in a formation of a contact binary ({\it bottom}).}
\end{figure}

If AM is conserved, the equilibrium radius of star B is slightly smaller
after the end of mass transfer than its Roche lobe (Fig.~3 top). As a
result, a semi-detached binary of the type of a short-period Algol is
formed, similar to W~Crv (Rucinski and Lu 2000). Such systems will exist
in a semi-detached state until the presently more massive star (star B
in our case) fills its Roche lobe due to radius increase and/or AML, and
a contact binary is formed. However, as modeling of mass exchange in
Algols indicate, a significant percentage of AM is lost during the mass
transfer in a common envelope phase, together with a small amount of
mass (Sarna and De Greve 1996), \eg for S~Cnc the results suggest 5\%
mass loss and 43\% AML. We can calculate how much AM our binary must
lose during Phase II to emerge as a contact system. The result shows
that the loss of 15\% of AM is enough for the equilibrium radius of star
B to become equal to the size of its Roche lobe and for the formation of
a contact binary consisting of two stars in thermal equilibrium (Fig.~3,
bottom). Such AML, accompanied by a mass loss of 3\%, was adopted when
plotting Fig.~2. The duration of Phase~II was assumed to be equal to
$10^8$ years which corresponds approximately to the thermal time scale.

At the end of Phase~II the components of the binary have $R_{\rm
A}=0.80$~\RS, $R_{\rm B}=1.02$~\RS and $q\approx0.51$. When plotted in the
mass--radius diagram, both components land in the regions occupied by
components of W-type contact binaries (see the starting points of the heavy
lines with arrows in Fig.~1).

\subsection{Phase III}
After the mass transfer on a thermal time scale has been completed the
system becomes a typical W-type contact binary in a global thermal
equilibrium. Large scale circulations in the common envelope carry
energy from star B to star A, which results in equalization of surface
temperatures of both stars (Martin and Davey 1995). K\"ahler (2004)
developed recently a model of a cool contact system in thermal
equilibrium. In his model the energy is transported from the hotter
component to the outermost part of the envelope of the cooler component,
separated from the bulk of the convective zone by a radiative layer. As
a result, deep convective layers of both components have different
entropies. In addition to circulations in the common envelope he also
assumed the existence of circulations in a radiative interior of the
less massive component, carrying a significant fraction of its nuclear
energy flux. The value of this fraction is an additional free parameter
of the problem. This removes the over-constraint in treatment of contact
binaries with a common convective envelope and permits to obtain a
unique solution in thermal equilibrium with the secondary component
adequately oversized. He applied this model to unevolved, or only
slightly evolved binaries. It would be interesting to extend his model
to the binary discussed in the present paper.

The analysis of the evolutionary models of low mass stars obtained by
Sien\-kie\-wicz shows that a helium core increases in these stars at an
approximate rate of 0.01~\MS/Gyr hence building a small (less than
0.1~\MS) helium core takes several Gyr. The mass--radius relations for
such stars are shown in Fig.~4. Note that models were evolved at a
constant mass (along the vertical lines in Fig.~4). The stellar radius
increases with the increasing mass of the helium core at a constant
stellar mass but it decreases with the decreasing stellar mass at a
constant core mass.
\begin{figure}[htb]
\vglue-3mm
\centerline{\includegraphics[width=11.9cm]{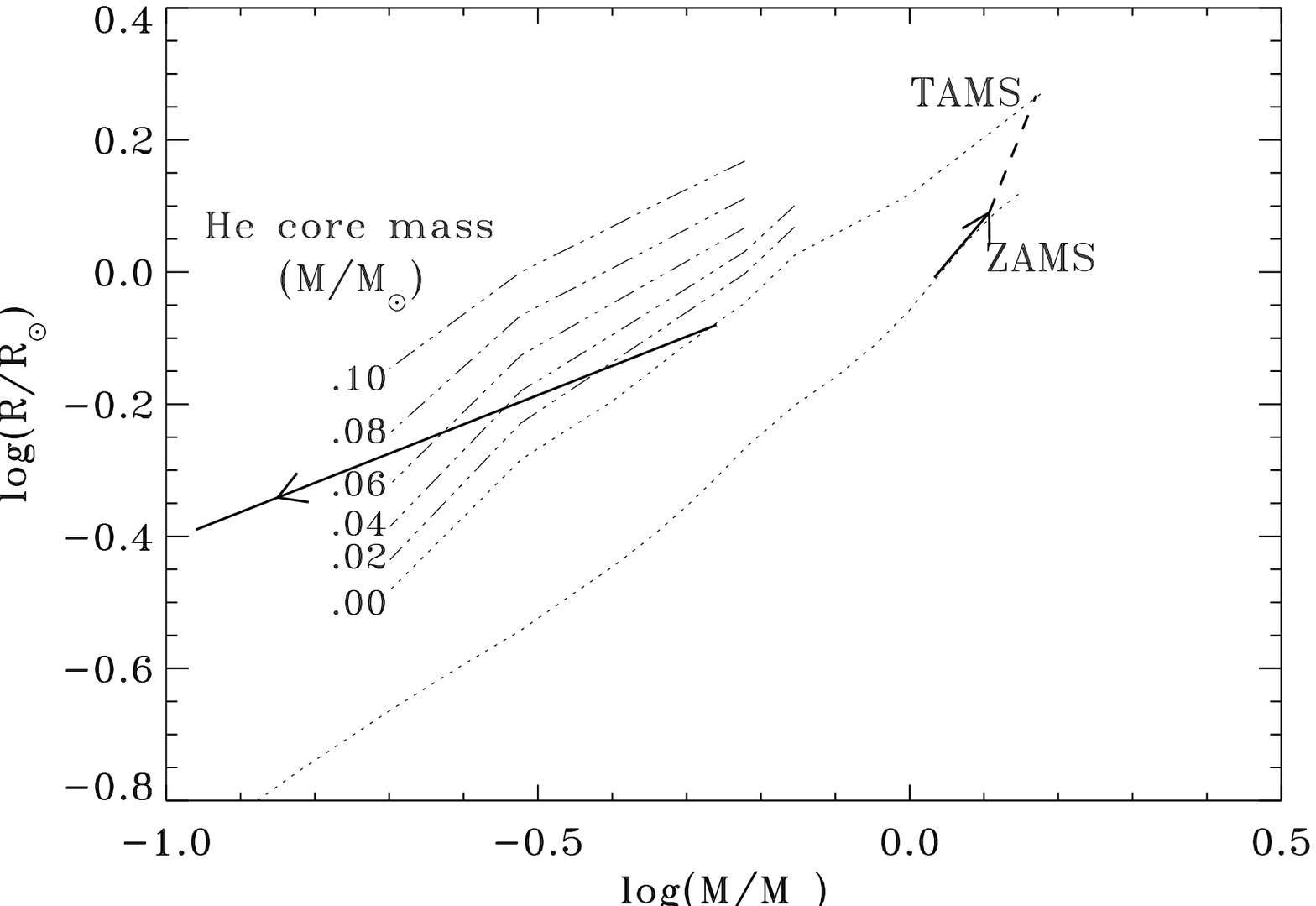}}
\vskip4pt
\FigCap{The evolution of the discussed binary during Phase~III. Dotted
lines show mass-radius relations for ZAMS and TAMS stars and dash-dotted
lines show the same relations for stars with different helium cores in
solar mass, as indicated. Heavy solid and broken lines are schematic
evolutionary tracks of both components of the discussed binary, with
arrows indicating the direction of evolution.}
\end{figure}

Assuming that the AML rate during Phase~III has its saturated value
found for Phase~I we obtain again a coincidence of both time scales
governing the evolution in contact: nuclear time scale of the lower mass
component and orbital AML time scale. As a result, it is possible to
calculate approximate evolutionary tracks of both components in the
mass--radius diagram (Fig.~4, see also Fig.~1) and the time variations of
the orbit parameters (Fig.~2). After about 4~Gyr the mass of star A
decreases to about 0.3~\MS whereas the mass of its helium core increases
to 0.04~\MS. Average net mass transfer from star A to B during this time
is $6\times10^{-11}$~\MS/year. This is hydrogen rich matter. Because
star B converts in thermonuclear reactions about $10^{-11}$~\MS/year of
hydrogen into helium, which is several times less, its degree of
chemical non uniformity decreases and it stays close to ZAMS. When
its mass increases beyond 1.3~\MS, the convection zone gradually
disappears, which is accompanied by a decrease of magnetic activity. The
system evolves into A-type contact binary with the primary star
unevolved. However, when the mass of star A becomes significantly less
than the initial core mass with nuclear reactions, \ie 0.25--0.30~\MS,
helium enriched matter starts to flow to star B. At the same time its
rate of conversion hydrogen to helium increases due to increased
luminosity. Both these effects increase the chemical non uniformity of
star B so it moves away from ZAMS. After another 1 Gyr the mass of star
A decreases beyond 0.1~\MS and both components merge into a single,
rapidly rotating star (Rasio 1995). This completes Phase~III.

\Section{Discussion and Conclusions}
\vspace*{-6pt}
\subsection{Uncertainties}
Because of a significant uncertainty of the AML rate, the time needed to
reach the Roche lobe by a more massive component in Phase~I is known with
accuracy of only $\approx50\%$ (see Eq.~1). In our scenario it was 
assumed that the star fills its Roche lobe just reaching TAMS. With AML
rate decreased to 50\% of the adopted value star A reaches TAMS while still
being inside the Roche lobe and an additional 1~Gyr is needed to fill the
lobe whereas for AML rate increased by 50\% star A reaches its critical
lobe when it is close, but not yet exactly at TAMS. This is a minor
modification of the further evolution of the discussed system.

The adopted mass loss rate is known not better than to a factor of 2 (Wood
\etal 2002) but because it is low anyway, it has a very small influence on
the considered time scale. Only for values several times higher than
adopted, the mass loss will significantly modify the evolution of a close
binary (Eggleton and Kiseleva-Eggleton 2002, Yakut and Eggleton 2005).

A good example of a detached binary in this phase is a very active system
XY~UMa with the orbital period of 0.48 of a day. According to Pribulla
\etal (2001) the components have masses of 1.1~\MS and 0.66~\MS, and radii
of 1.16~\RS and 0.63~\RS, respectively. The primary fills already 68\% of
its Roche lobe whereas the secondary fills only 22\% of its Roche lobe. In
about 1--2~Gyr the primary (\ie star A in our notation) fills up its Roche
lobe and mass transfer begins.

The beginning of mass transfer initiates the second phase. It has been
usually accepted that after accreting about 0.1~\MS the less massive
star swells, fills its Roche lobe and a contact binary is formed with a
common convective envelope within which a strong energy flux flows
through the neck, keeping the less massive star oversized (Lucy 1968,
1976, Flannery 1976, Webbink 1976, Sarna and Fedorova 1989). Recent
observations of several short period Algols with periods down to 0.4 of
a day contradict, however, this picture (Shaw 1994, Rucinski and Lu
2000, Pribulla \etal 2003). Such stars must have gone through mass
exchange with the mass ratio reversal just like longer period ordinary
Algols. We are still unable to model correctly the common envelope phase
of mass exchange with all hydrodynamical effects in 3D geometry, energy
balance, etc. taken properly into account. It is therefore adopted in
the present paper that mass exchange, taking place in the discussed
close binary, ends with mass ratio reversal. The variable V361~Lyr with
orbital period of 0.31 of a day, recently discussed by Hilditch \etal
(1997) seems to be now in the second phase. According to their analysis
the more massive component (1.26~\MS) fills its Roche lobe and transfers
mass at a rate of $10^{-7}$~\MS/yr to the secondary (0.87~\MS) which has
a radius equal to only 50\% of its Roche lobe. A hot spot is visible at
the location where the mass stream falls on the surface of the
secondary. Using the results of Sarna and Fedorova (1989) the authors
predict that mass transfer will last mere $\approx10^6$ years and will
end with swelling of the secondary, until it fills its Roche lobe. They
stress that V361~Lyr is a very exceptional system caught in an extremely
short lasting phase of evolution. We should note, however, that this
interpretation requires even more exceptional situation. The secondary
does not yet show any sign of expansion. It means that the star is
either caught at the very beginning of the mass transfer phase when it
has not yet had time to swell upon accreting mass or that the secondary
can accommodate the accreted matter without substantial swelling.

In case of a binary with mass ratio distinctly different from one, as
considered here, a conservative mass exchange results in a formation of
the short-period Algol. The binary spends the first part of Phase~III in
a semi-detached configuration until a contact system is formed. For a
binary with an initial mass ratio $q\approx1.2$ a conservative mass
exchange in Phase~II takes place on a nuclear time-scale because any
mass loss from star A results in shrinkage of its radius below the size
of the Roche lobe. Further mass loss requires an evolutionary expansion
of star A (St{\c e}pie{\'n} 2006). The binary spends a noticeable time in a
near contact phase, with a more massive component losing mass on a
nuclear time scale. After mass ratio reversal an ordinary Algol is
formed again. The duration of the Algol evolutionary stage can be
shortened, or skipped altogether if enough AM is lost during Phase~II.
In any case a contact binary is ultimately formed with both components
in thermal equilibrium. Later, it evolves towards an extreme mass ratio
A-type binary. The common envelope, in which transport of energy takes
place, is rather thin (a fraction of mass lying above the critical Roche
surface) as discussed recently by K\"ahler (2004) and it is separated
from the bulk of the convective zone of star A by a radiative layer. The
separation results in different entropies of deep convective envelopes
of both stars. The star W~Crv, mentioned above, is in the Algol phase
(Rucinski and Lu 2000) whereas $\epsilon$~CrA, an A-type contact binary
with $q=0.13$, is in a more advanced stage of the third phase. Stars
like SX~Crv ($q=0.07$) or AW~UMa ($q=0.08$) are probably close to
coalescence.

The rate of evolution of the binary in Phase~III is determined by the AML
rate and evolutionary effects of both components. These processes result in
a net mass transfer from star A to star B until coalescence occurs.
Unfortunately, as the following discussion shows, the total AML rate in
Phase~III is known with even greater uncertainty than that indicated by
Eq.~(1). The binary spends several (five in our case) Gyr in this
phase. The time spent in Phase III may be sufficient for star B to complete
its main sequence evolution. Its subsequent expansion will cause the common
envelope to approach the outer critical Roche surface until a mass loss
from the critical point L2 occurs. As the detailed calculations show, the
specific AM of the matter flowing from L2 to infinity is 10--20 times
larger than the specific orbital AM (Shu, Lubow and Anderson 1979). It
means that the mass loss from L2 is as efficient in removing AM from the
binary as the magnetized wind. For a significant mass loss from L2 the
total AML rate may be considerably increased over that resulting solely
from the wind. On the other hand, AML rate {\it via} a magnetized wind may
be substantially lower in contact binaries than in detached
systems. Measured X-ray fluxes of W~UMa type variables are lower by a
factor of 4--5 compared to rapidly rotating single stars (St{\c e}pie{\'n},
Schmitt and Voges 2001). If AML rate scales with the X-ray flux we expect
the correspondingly lower AML rates in contact binaries.  The unknown
amount of AML from L2 as well as of AML rate {\it via} a wind increase
substantially an uncertainty of the adopted AML rate in Phase~III.

The luminosity of a binary in Phase III essentially does not differ from
a model with TRO. The total luminosity is dominated by the nuclear
luminosity of the more massive star which in both models is a MS object.
Depending on the degree of hydrogen depletion in the core its luminosity
changes approximately by a factor of two between ZAMS and TAMS (at a
constant mass). The contribution of the less massive component is
slightly larger in the present model than in the TRO model, because the
star burns hydrogen in the shell during Phase~III but the total
luminosity is changed insignificantly.

Fig.~5 shows the orbital period variations of the discussed system
through all three phases of binary evolution shown in Fig.~2. The shape
of the plotted curve suggests a separation between the period
distributions of detached and contact binaries. We expect the detached
binaries with solar type components to have periods longer than about
0.5 of a day whereas orbital periods of the contact binaries should be
concentrated around 0.3--0.4 of a day. Investigation of a large sample
of contact binaries from OGLE I program shows indeed a strong peak in
the period distribution of these stars at $P_{\rm orb}=0.35$ of a day
(Rucinski 1998). A similar peak was obtained by Szyma\'nski \etal
(2001). However, if AML rate is substantially reduced, the mass transfer
from star A to B, forced by an evolutionary expansion of star A, will
result in an increase of the orbital period when $q\rightarrow0$. This
should be visible as a systematic difference between periods of
variables with moderate and very low values of $q$ (\ie entering and
leaving Phase~III). Observations do not show this effect. The average
orbital period of 11 binaries with $q<0.15$, listed in the catalog by
Pribulla \etal (2003), is equal to 0.4 of a day. Obviously, a
self-regulating mechanism takes place: a decrease of AML rate, with the
mass flux unchanged, results in an increase of the period, hence sizes
of the Roche lobes, which, in turn, lowers the mass flux until the
period increase is stopped. Lower AML rate should result in the
lengthening of a duration of Phase~III. For a significantly higher AML
rate than adopted here (\eg due to intense mass loss from L2) the
duration of Phase~III will be correspondingly shortened.
\begin{figure}[htb]
\centerline{\includegraphics[width=11.6cm]{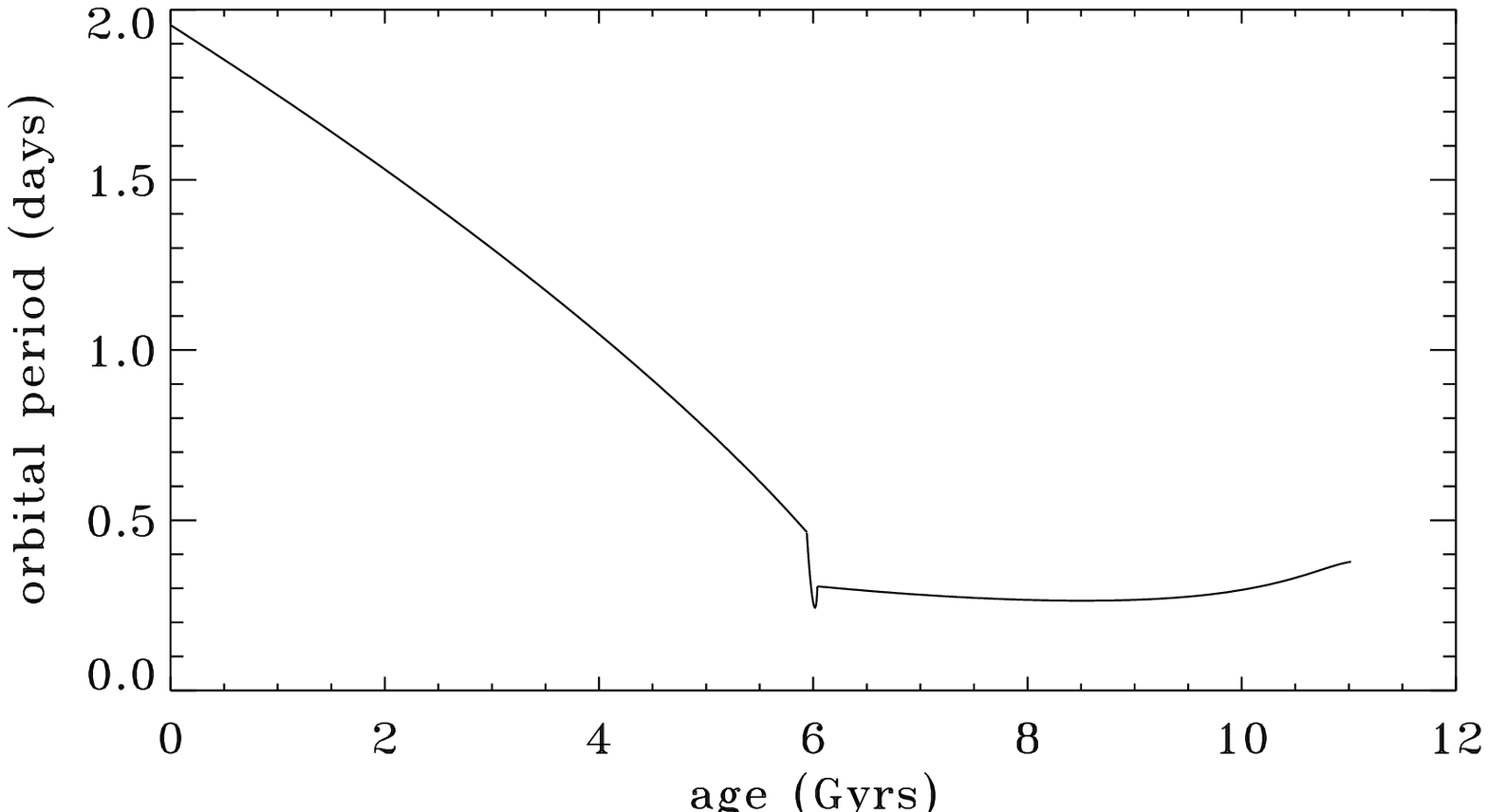}}
\vskip4pt
\FigCap{Time behavior of the orbital period of the discussed binary
through all three phases.}
\end{figure}

\subsection{Period Variations}
Many cool contact binaries exhibit period variations (\eg Kreiner, Kim
and Nha 2001, Qian 2003). While many $O-C$ curves are flat or show only
quasi periodic variations, parabolas have been fitted to others,
suggesting systematic period increase/decrease. Because a typical time
interval covered by the observations extends over only some tens of
years, the resulting detection threshold for period variations is
$\dot{P}\approx10^{-7}$. Several stars show period variations of this
order (Qian 2002), which gives the corresponding time scale equal
roughly to the stellar thermal time scale. Some authors interpret these
variations as resulting from mass transfer between the components. Such
alternate mass transfer (visible as a respective period
increase/decrease) has been predicted by the TRO theory (Lucy 1976,
Flannery 1976) and the observations are interpreted in support of this
model.

The recent, precise observations of a few hundred of W~UMa type stars by
the OGLE team, with a detection threshold for period variations of
$10^{-8}$, showed that most of the observed stars reveal period variations
of this order or less. Their distribution can be fitted with a Gaussian
centered at zero (Kubiak, private communication). Variations exceeding
substantially the threshold value are rare. Such a distribution is at odds
with the TRO theory. If a large fraction of W~UMa type stars were in the
TRO phase, we would expect a two point distribution centered roughly at
$\pm(10^{-7}{-}10^{-8})$. The Gaussian distribution is expected for random
period variations. Random variations can arise \eg from fluctuations in
mass transfer associated with the energy transfer between the
components. To transfer the required amount of energy (of the order of
0.5~\LS) from the hotter to cooler component in a convective way, of the
order of $2\times10^{-5}$~\MS/year must be transferred (back and
forth) through the neck between the stars. The rough estimate of energy
transfer along the equipotential surface, \ie at a constant pressure is
$$\Delta L=c_p\dot M\Delta T\eqno(7)$$
where $\Delta L$ is the transferred luminosity, $c_p$ specific heat at the
constant pressure, $\dot M$ mass flux and $\Delta T$ is the difference in
temperature between the matter flowing from the hotter to the cooler
component and returning back.

As a conservative estimate, $2\times10^8$~erg/g$\cdot$K was adopted for
the value of $c_p$. This is the largest value appearing in tables of
convective envelopes given by Baker and Temesvary (1966). A rather high
value of $10^3$~K was adopted for $\Delta T$. For lower $c_p$ and/or
less cooling, the mass flux will be correspondingly higher (Webbink 1977
estimates this flux at $6\times10^{-4}$~\MS/year). The typical observed
period variations require the mass transfer of the order of
$10^{-8}$~\MS/year or less, which is only $\approx10^{-3}{-}10^{-4}$ of
the estimated mass flux. Such fluctuations can be expected, \eg as a
result of magnetic activity variations. Significant change of coverage
of the stellar surface by dark spots perturbs the thermal structure of
the stellar convection zone and it may result in a slight change of the
stellar radius on a thermal time scale (Spruit 1982) modifying
correspondingly the mass flux.

\subsection{Summary of the Main Conclusions}
An evolutionary scenario is presented for formation and evolution of cool
contact binaries of W~UMa type. W~UMa type binaries are defined as binaries
in which the initial masses of both components do not exceed the mass limit
for the existence of a subphotospheric convection zone. It is assumed that
detached, cool binaries with short orbital periods (about 2--3 days) and
solar type primaries (\ie with masses $\approx0.9{-}1.3$~\MS) are progenitors
of contact binaries. The assumption is in agreement with the current view
about the origin of close binaries. The numerical models seem to exclude
fission and they favor early fragmentation of the protostellar cloud as the
way of close binary formation, with the result that ZAMS binaries are
expected to have orbital periods not shorter than about 2 days.
Observations of very young stars support this view.

The binaries lose angular momentum (AM) {\it via} magnetized wind from
both components. As it turns out, the time scale for orbital evolution
due to AML of such binaries is equal to several Gyr, or more, depending
(primarily) on the length of the initial orbital period. Such time
scales are the same as nuclear time scales for MS evolution of
primaries. As a result of this coincidence, the primary is evolutionary
advanced (\ie it is close to, or beyond TAMS) when the orbit shrinks,
due to AML, so that the critical Roche lobe approaches its surface.
According to this scenario we should observe few, if any, young
W~UMa-type stars. Observations support this picture: the number of such
stars is very low in stellar clusters with ages less than about
4--4.5~Gyr and it increases rapidly in older ones.

The Roche lobe overflow results in mass transfer to the secondary. It is
assumed in the present scenario that the mass transfer continues
(through the common envelope phase) until the mass ratio reversal. Such
an exchange mass episode is similar to the one taking place in Algols
and it is different from that adopted by Lucy (1976) and Flannery (1976)
in case of W~UMa stars. They assumed that specific entropy of the
convective envelopes of both components takes the same value during the
common envelope phase. As a result, net mass exchange stops after
transferring a modest amount of matter and is replaced by energy
transfer. The turbulent convective energy transport between the
components keeps both convective zones on the same adiabatic curve but
the resulting configuration of both stars is in global thermal
nonequilibrium. Thermal Relaxation Oscillations are necessary for a long
time scale existence of such binaries. However, the recent models of
energy transfer between the components of a W~UMa type binary indicate
that large scale circulations, flowing in a relatively thin common
envelope (above the inner critical Roche surface), can transport energy
with a required efficiency (Martin and Davey 1995, K\"ahler 2002a,
2002b, 2004), hence the requirement of identical specific entropy in
both convective layers is no longer needed. This favors the Algol-type
mass transfer with mass ratio reversal. The existence of several Algols
with periods close to 0.5 of a day supports the scenario. The system
emerges as a contact binary, if enough AM is lost during the common
envelope or as a short-period Algol which transforms, after losing excess
AM, also in a contact binary.

The consecutive evolution in contact is governed by AML, assumed to be
as efficient as in the detached phase, and the nuclear evolution of the
present secondary which is the formerly more massive component. The
secondary is hydrogen depleted in the center and it builds a small
helium core. Time scales of both processes are again close to one
another. AML keeps the orbit compact whereas nuclear evolution forces
secular mass transfer from the secondary to the primary. The binary
evolves towards an extreme mass ratio system, which, after several Gyr
ends with a merging of both components.

\Acknow{I thank Dr.\ Ryszard Sienkiewicz for calculating several
evolutionary models of low mass stars. My thanks are also due to
Professors Stefan Mochnacki, Bohdan Paczynski and Slavek Rucinski for
reading and commenting on the manuscript. Special thanks are due to
Professor Wojciech Dziembowski, the referee, whose thorough report
helped to improve substantially the text of the paper. This work was
partly supported by KBN grant 1 P03 016 28.}

\end{document}